# Top quark pair production Cross Section in ATLAS with early data

Babak Abi, for the ATLAS Collaboration,
*Oklahoma State University, OK 74078, USA*

A search is performed for top quark pairs with early data of ATLAS pp collision taken at √s = 7 TeV. Several candidate events are observed, in both lepton plus jets and dilepton channels. The properties of these events are described, and compared to the expectations from Monte Carlo simulation. A first study of the backgrounds in the lepton plus jets channel is presented including a data-driven determination of the contribution of QCD multi-jet events.

## 1. INTRODUCTION

The top quark plays an important role in our understanding of the fundamental world and their studies are a crucial component of the ATLAS physics program. LHC is a top factory, we expect the production of one top quark pair decaying into the $e/\mu$ + jets channel for every ~20 nb$^{-1}$ of recorded LHC data, as well as one top quark pair decaying into the $ee/e\mu/\mu\mu$ + jets channel for every ~110 nb$^{-1}$ of data. While event selection efficiency will reduce the number of observable candidates, a handful of such candidates are expected to be selected with O(300) nb$^{-1}$ of data.

This paper presents observed candidates in the first 280 nb$^{-1}$ of ATLAS pp collision data taken at √s = 7 TeV, using an event selection designed for an early measurement of the $t\bar{t}$ production cross section. After introducing the data selection and corresponding samples of Monte Carlo simulated data, the object and event selections are described, the properties of selected candidate events are discussed, and compared with the expectation from simulation. Finally the results from study of the backgrounds in the lepton plus jets channel is presented with QCD multi-jet contributions predicted from data by the so-called 'matrix method'.

## 2. DATA AND SIMULATED EVENTSAMPLE

The ATLAS detector [1] covers nearly the entire solid angle around the collision point with layers of tracking detectors, calorimeters and muon chambers. Top event candidates are reconstructed using all these detector from the data recorded until 19th July 2010 (~ 280 nb$^{-1}$ at √s = 7 TeV ). This luminosity estimate has an uncertainty of 11% [3].

For the generation of $t\bar{t}$ signal and single top samples, MC@NLO v3.41 was used, with PDF set CTEQ66, assuming a top mass of 172.5 GeV [2]. For the generation of W+($b\bar{b}$)+jets and Z+($b\bar{b}$)+jets, ALPGEN v2.13 was used, All events were hadronized with HERWIG, using Jimmy for the underlying event model [2]. Subsequent detector and trigger simulation, followed by offline reconstruction, has been performed with standard ATLAS software making use of GEANT4. The effect of pileup is expected to be small for the studies presented at this paper.

## 3. EVENT SELECTIONS

The selections for both lepton+jets and dilepton events start by requiring the presence of at least one lepton ($e$ or $\mu$) associated with a leptonic high-level trigger object of the same type with a 10 GeV transverse momentum ($p_T$) threshold. Events must have a reconstructed primary vertex with at least 5 tracks, and are discarded if any jet with $p_T$ > 10GeV at the electro-megnatic (EM) scale fails jet quality cuts designed to reject jets arising from out-of-time activity or calorimeter noise. These quality cuts remove a negligible fraction of simulated events.

The dilepton selection requires two oppositely-charged leptons (*ee, μμ* or *eμ*) each satisfying $p_T > 20$ GeV, at least one of which must be associated to a leptonic high-level trigger object. At least two jets with $p_T > 20$ GeV are required, but no b-tagging requirements are imposed. In the *ee* channel, to suppress backgrounds from Drell-Yan and QCD multi-jet events, the missing transverse energy must satisfy $E_T^{MISS} > 40$ GeV, and the invariant mass of the two leptons must be at least 5 GeV from the Z-boson mass, *i.e.* $|M_{ee} - M_Z| > 5$ GeV. For the muon channel, the corresponding requirements are $E_T^{MISS} > 30$ GeV and $|M_{\mu\mu} - M_Z| > 10$ GeV. For the *eμ* channel, where the background from Z → *ee* and Z → *μμ* is expected to be much smaller, no $E_T^{MISS}$ or Z-mass veto cuts are applied, but the event HT, defined as the scalar sum of the transverse energies of the two leptons and all the selected jets, must satisfy $H_T > 150$ GeV.

The lepton+jets selection then requires the presence of exactly one offline-reconstructed electron or muon with $p_T > 20$ GeV, satisfying the object requirements described in [3] and matching a leptonic high-level trigger object within $\Delta R < 0.15$. At least 4 jets (selected as described in [3]) with $p_T > 20$ GeV and pseudo-rapidity $|\eta| < 2.5$ are then required, at least one of which must be *b*-tagged. Finally, the missing transverse energy must satisfy $E_T^{MISS} > 20$ GeV.

## 4. CANDIDATE EVENTS AND DISTRIBUTIONS

Tables 1 and 2 show the events observed in data passing the all event selection criteria in the lepton+jets and dilepton channels, and list some of the basic properties of these events. Table 1 lists for the lepton plus jets candidates, the transverse mass $m_T$ of the lepton and $E_T^{MISS}$ and three-jet mass $m_{jjj}$ that is defined as the invariant mass of the three-jet combination with the largest $p_T$. Table 2 lists the dilepton candidates with also energy sum $H_T$. Some of the event properties are indicated as arrows in the corresponding distributions in Fig.1a to Fig.1d.

## 5. ESTIMATE OF QCD MULTI-JET BACKGROUND

The QCD multi-jet background is calculated from the data using the Matrix Method (MM) [4]. MM is based on making two set of samples: the 'tight' sample passing all event selection cuts defined above, and a 'loose' sample where the lepton identification cuts are relaxed. Then, using following formula one can estimate the QCD multi-jet background;

$$N^{loose} = N_{real}^{loose} + N_{fake}^{loose}, \quad \epsilon_{fake} = N_{fake}^{tight}/N_{fake}^{loose}$$
$$N^{tight} = \epsilon_{real}N_{real}^{loose} + \epsilon_{fake}N_{fake}^{loose}, \quad \epsilon_{real} = N_{real}^{tight}/N_{real}^{loose}$$

The $\epsilon_{real}$ is taken from Monte Carlo simulated Z → *ll* events and $\epsilon_{fake}$ is taken from a real data control sample enriched in QCD multijet events, selected by requiring at least one jet with $p_T > 20$ GeV and $E_T^{MISS} < 10$ GeV. The control sample has a small residual contamination from W and Z events and this procedure has to be iterated. The MM procedure is performed in bins to yield estimates as a function of jet multiplicity and pseudorapidity ($\eta$) of the electron. A result of this study without the b-tagging requirement is shown in fig 2a (2b) and without *b*-tagged requirement shown in fig 2c (2d).

## 6. CONCLUSION

A search for top quark candidates has been conducted with 280 nb$^{-1}$. Strategy has been defined with specific selections designed for an early measurement of the cross-section in lepton+jets and dilepton channels. Nine candidate events were seen: 4 *e*+jets, 3 *μ*+jets candidates, 1 *ee*, and 1 *eμ* candidates. Event counts are consistent with expectations from simulation.

| ID | Run number | Event number | Channel | $p_T^{lep}$ (GeV) | $E_T^{miss}$ (GeV) | $m_T$ (GeV) | $m_{jjj}$ (GeV) | #jets $p_T > 20$ GeV | #*b*-tagged jets |
|---|---|---|---|---|---|---|---|---|---|
| LJ1 | 158801 | 4645054 | *μ*+jets | 42.9 | 25.1 | 59.3 | 314 | 7 | 1 |
| LJ2 | 158975 | 21437359 | *e*+jets | 41.4 | 89.3 | 68.7 | 106 | 4 | 1 |
| LJ3 | 159086 | 12916278 | *e*+jets | 26.2 | 46.1 | 62.6 | 94 | 4 | 1 |
| LJ4 | 159086 | 60469005 | *e*+jets | 39.1 | 66.7 | 102 | 231 | 4 | 1 |
| LJ5 | 159086 | 64558586 | *e*+jets | 79.3 | 43.4 | 86.7 | 122 | 4 | 1 |
| LJ6 | 159224 | 13396261 | *μ*+jets | 29.4 | 65.4 | 64.1 | 126 | 5 | 1 |
| LJ7 | 159224 | 13560451 | *μ*+jets | 78.7 | 40.0 | 83.7 | 108 | 4 | 1 |

Table 1 List of candidate $t\bar{t}$ events in the lepton+jets channel in data,

| ID | Run number | Event number | Channel | $p_T^{lep}$ (GeV) | $E_T^{miss}$ (GeV) | $H_T$ (GeV) | #jets $p_T > 20$ GeV | #*b*-tagged jets |
|---|---|---|---|---|---|---|---|---|
| DL1 | 155678 | 13304729 | *ee* | 55.2/40.6 | 42.4 | 271 | 3 | 1 |
| DL2 | 158582 | 27400066 | *eμ* | 22.7/47.8 | 76.9 | 196 | 3 | 1 |

Table 2 List of candidate $t\bar{t}$ events in the dilepton+jets channel in data,

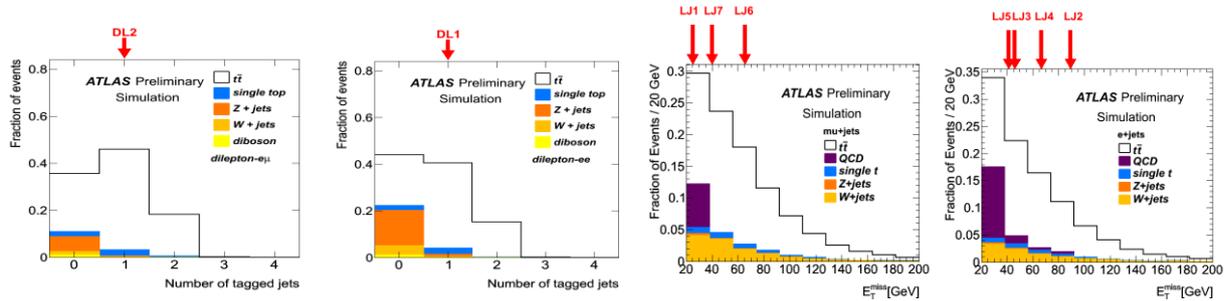

Figure 1 Multiplicity distributions for b-tagged jets for dilepton, *ee*(DL1) and *eμ*(DL2) events in the two left hand plots and $E_T^{MISS}$ for lepton plus jets events in the two right hand plots.

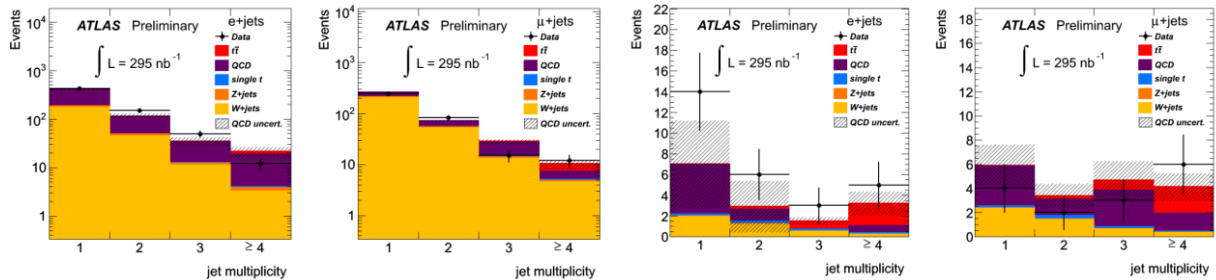

Figure 2 Jet multiplicity distributions without (left) and with (right) one *b*-tagged jet for lepton plus jets.